\documentclass[a4paper]{article}

\usepackage[ninepoint]{itgspeech2025}
\usepackage{times}            %% Choose Times Roman font
\usepackage[english]{babel}   %% This is an English paper
\usepackage[utf8]{inputenc}
\usepackage[T1]{fontenc}      %% use PS Type1 fonts
\usepackage[sort&compress,numbers]{natbib}	%% bibliography
\usepackage{amsmath,amssymb}
\usepackage{graphicx}
\usepackage[colorlinks=false,pdfborder={0 0 0}]{hyperref}
\usepackage{units}
\usepackage{enumitem}
\usepackage{makecell}
\usepackage{multirow}
\usepackage{booktabs}
\usepackage[caption=false]{subfig} % Allows subfigures
\usepackage{svg}
\usepackage{caption}
\usepackage{bm}

\title{Investigation of Speech and Noise Latent Representations in Single-channel VAE-based Speech Enhancement}

\author{Jiatong Li, Simon Doclo}

\address{Dept. of Medical Physics and Acoustics and Cluster of Excellence Hearing4all, Carl von Ossietzky Universität Oldenburg, Germany\\
  Email: \texttt{\{jiatong.li,simon.doclo\}@uni-oldenburg.de}}

\begin{document}

\maketitle
\begingroup
\renewcommand\thefootnote{}\footnotetext{This work was funded by the Deutsche Forschungsgemeinschaft (DFG, German Research Foundation) under Germany’s Excellence Strategy - EXC 2177/1 - Project ID 390895286 and Project ID 352015383 - SFB 1330 B2.}
\endgroup

\begin{abstract}
%% Place your abstract here
Recently, a variational autoencoder (VAE)-based single-channel speech enhancement system using Bayesian permutation training has been proposed, which uses two pretrained VAEs to obtain latent representations for speech and noise. Based on these pretrained VAEs, a noisy VAE learns to generate speech and noise latent representations from noisy speech for speech enhancement. Modifying the pretrained VAE loss terms affects the pretrained speech and noise latent representations. In this paper, we investigate how these different representations affect speech enhancement performance.  Experiments on the DNS3, WSJ0-QUT, and VoiceBank-DEMAND datasets show that a latent space where speech and noise representations are clearly separated significantly improves performance over standard VAEs, which produce overlapping speech and noise representations.
\end{abstract}

%% And now start with your paper content

%%%%%%%%%%%%%%%%%%%%%%%%%%%%%%%%%%%%%%%%%%%%%%%%%%%%%%%%%%%%%%%%%%
\section{Introduction}
%%%%%%%%%%%%%%%%%%%%%%%%%%%%%%%%%%%%%%%%%%%%%%%%%%%%%%%%%%%%%%%%%%
The goal of speech enhancement is to remove background noise from noisy speech signals, thereby improving speech intelligibility and speech quality. 
Recently, several generative models, e.g. based on the variational autoencoders (VAEs)\cite{9414060,bie2022unsupervised,sadeghi2023fast,10446294,10447837,xiang2022bayesian,xiang2022deep,xiang2023two,xiang2024deep}, generative adversarial networks\cite{pascual17_interspeech,baby2019sergan,phan2020improving,wali2022generative,10508391}, and diffusion models\cite{lu2022conditional,richter2023speech,nortier2024unsupervised,gonzalez2024investigating}, have been proposed for speech enhancement. 
VAEs, consisting of an encoder and a decoder, aim to model high-dimensional data distributions and are typically trained by maximizing the evidence lower bound (ELBO)\cite{Kingma2014}. The encoder extracts the information of the input data into latent representations, while the decoder reconstructs the data from the latent representations. VAEs have been used in several speech processing tasks to learn interpretable latent representations. For example, in voice conversion, VAEs have been used to disentangle spoken content and speaker identity\cite{hsu18_interspeech,lian2022robust,lu2023disentangled}, while in speech synthesis, VAEs have been used to disentangle prosody, content, acoustic details and timbre\cite{naturalspeech}.

\begin{figure}[t!]
    \centering
    \includegraphics[width=0.4\textwidth]{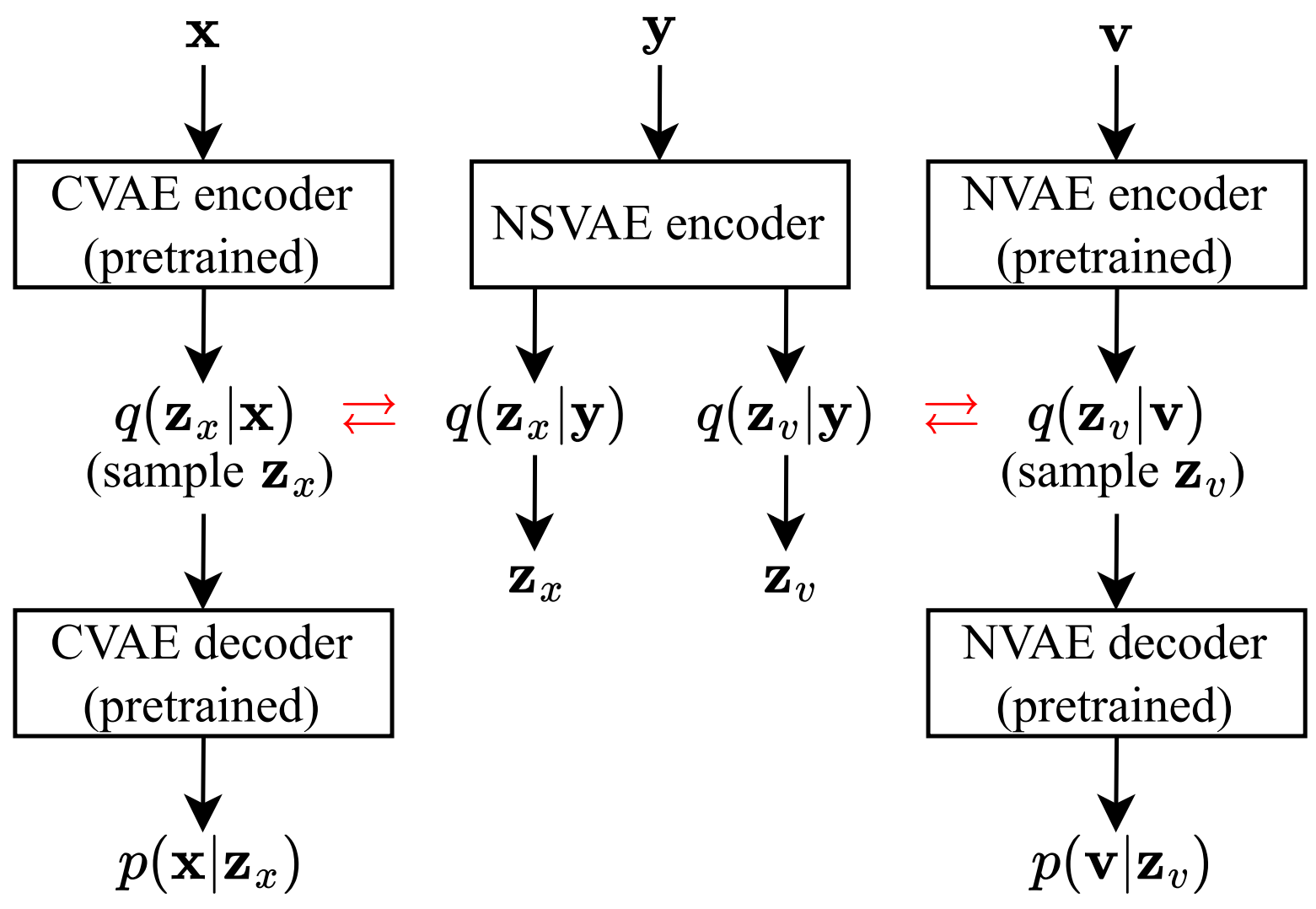} % Include the SVG image
    \caption{Overview of the PVAE system, consisting of two pretrained VAEs, i.e. clean speech VAE (CVAE) and noise VAE (NVAE), and the noisy VAE (NSVAE).} % Add a caption
    \label{fig:system diagram} % Optional: Add a label for referencing
\end{figure}

Since the VAE framework facilitates efficient posterior inference and reliable reconstruction, several VAE-based approaches have been proposed for single-channel speech enhancement. In \cite{bie2022unsupervised,sadeghi2023fast,10446294,10447837,9414060}, VAEs have been combined with non-negative matrix factorization (NMF) to model speech and noise. Since NMF may limit performance compared to deep learning models, in \cite{xiang2022bayesian, xiang2022deep} a Bayesian permutation training (PVAE) speech enhancement system has been introduced (see Fig.\ref{fig:system diagram}), which uses two pretrained VAEs to extract latent representations for clean speech (CVAE) and noise (NVAE). A third VAE, noisy VAE (NSVAE), learns to generate latent clean speech and noise representations from a noisy speech by learning the latent speech and noise distributions of the pretrained CVAE and NVAE. In inference, the NSVAE encoder and the CVAE and NVAE decoders are then used to estimate clean speech from noisy speech.
In subsequent work, the performance of the PVAE system has been improved by using adversarial training \cite{xiang2023two} and by adopting the more advanced DCCRN network architecture \cite{xiang2024deep}.

In this paper, we consider the PVAE system from \cite{xiang2022deep} and focus on latent speech and noise representations from the pretrained CVAE and NVAE. Because the latent representations from the pretrained VAEs guide the behavior of the NSVAE encoder and are hence crucial in estimating clean speech, we investigate the influence of different pretrained speech and noise latent representations. Instead of using the standard VAE and ELBO loss as in \cite{xiang2022bayesian, xiang2022deep} for the pretrained VAEs, we propose to use the Disentangled Inferred Prior VAE (DIP VAE) \cite{kumar2018variational}. Since each term in the DIP VAE loss affects the latent representations, we investigate several modifications of the DIP VAE loss for the pretrained VAEs and explore how different latent representations influence speech enhancement performance. Through experiments on a matched dataset (DNS3) and two mismatched datasets (WSJ0-QUT, VoiceBank-DEMAND), we demonstrate that the latent representations from the pretrained VAEs significantly influence the speech enhancement performance of the PVAE system. Specifically, a latent space where pretrained clean speech and noise representations are clearly separated improves performance in both matched and mismatched datasets. In contrast, using the standard VAE as in \cite{xiang2022deep} for the pretrained VAEs turns out to be suboptimal, as it produces overlapping speech and noise representations in the latent space.

%%%%%%%%%%%%%%%%%%%%%%%%%%%%%%%%%%%%%%%%%%%%%%%%%%%%%%%%%%%%%%%%%%
\section{VAE-based Speech Enhancement}
\label{sec:paper_format_style}%
%%%%%%%%%%%%%%%%%%%%%%%%%%%%%%%%%%%%%%%%%%%%%%%%%%%%%%%%%%%%%%%%%%
In this section, the PVAE speech enhancement system from \cite{xiang2022deep} is reviewed. After presenting the signal model, the training of different VAEs and the inference process are described in more detail.

\subsection{Signal Model}
%%%%%%%%%%%%%%%%%%%%%%%%%%%%%%%%%%%%%%%%%%%%%%%%%%%%%%%%%%%%%%%%%%

The PVAE system performs speech enhancement in the short-time Fourier transform (STFT) domain. In the STFT domain, the observed noisy speech $\mathbf{Y}_n \in \mathbb{C}^F$ at the time frame $n \in [1, N]$, where $N$ and $F$ denote the number of time frames and frequency bins, is given by
\setlength{\abovedisplayskip}{6pt} % Space above equations
\setlength{\belowdisplayskip}{5pt} % Space below equations
\begin{equation*} 
    \mathbf{Y}_n = \mathbf{X}_n + \mathbf{V}_n, \tag{1}
\end{equation*}
where $\mathbf{X}_n \in \mathbb{C}^F$ and $\mathbf{V}_n \in \mathbb{C}^F$ denote clean speech and noise, respectively. The log-power spectrum (LPS) of the noisy speech is defined as
\begin{equation*} 
    \mathbf{y}_n = \operatorname{log}_{10}|\mathbf{Y}_n|^2, \tag{2}
\end{equation*}
where $|\cdot|$ denotes the magnitude (element-wise) and $\mathbf{x}_n$ and $\mathbf{v}_n$ are defined similarly. For simplicity,  the time frame index $n$ is omitted in the following description. 

In \cite{xiang2022deep}, $\mathbf{y}$ is assumed to be generated from a random process involving the speech latent representation $\mathbf{z}_x \in \mathbb{R}^L$ and the noise latent representation $\mathbf{z}_v \in \mathbb{R}^L$, where $L$ denotes the dimension of the latent representations. The prior distributions for the latent representations $\mathbf{z}_x$ and $\mathbf{z}_v$ are denoted as $p(\mathbf{z}_x)$ and $p(\mathbf{z}_v)$. %This random process can be described by $p(\mathbf{y}|\mathbf{z}_x,\mathbf{z}_v)$. %
The generation processes of $\mathbf{x}$ and $\mathbf{v}$ from $\mathbf{z}_x$ and $\mathbf{z}_v$ are described by the likelihoods $p_{\scriptstyle\theta_x}(\mathbf{x}|\mathbf{z}_x)$ and $p_{\scriptstyle \theta_v}(\mathbf{v}|\mathbf{z}_v)$, and $\mathbf{z}_x$ and $\mathbf{z}_v$ can be sampled from the speech and noise posterior distributions $q_{\scriptstyle \phi_x}(\mathbf{z}_x|\mathbf{x})$ and $q_{\scriptstyle \phi_v}(\mathbf{z}_v|\mathbf{v})$. Assuming that the above-mentioned distributions are estimated by VAEs, $\phi_x$ and $\theta_x$ denote the encoder and decoder parameters of the clean speech VAE (CVAE), while $\phi_v$ and $\theta_v$ denote the encoder and decoder parameters of the noise VAE (NVAE). Assuming $\mathbf{z}_x$ and $\mathbf{z}_v$ to be independent, $\mathbf{z}_x$ and $\mathbf{z}_v$ can also be sampled from the noisy posterior distribution, $q_{\scriptstyle \phi_y}(\mathbf{z}_x, \mathbf{z}_v|\mathbf{y})=q_{\scriptstyle \phi_y}(\mathbf{z}_x|\mathbf{y})q_{\scriptstyle \phi_y}(\mathbf{z}_v|\mathbf{y})$, where $\phi_y$ denotes the encoder parameters of the noisy VAE (NSVAE). In the following, the encoder and decoder parameters are omitted for simplicity.

\subsection{System Description}
%%%%%%%%%%%%%%%%%%%%%%%%%%%%%%%%%%%%%%%%%%%%%%%%%%%%%%%%%%%%%%%%%%

In the causal PVAE speech enhancement system (see Fig.\ref{fig:system diagram}), the CVAE and NVAE are pretrained models that extract latent representations for clean speech $\mathbf{x}$ and noise $\mathbf{v}$. The NSVAE aims at disentangling speech and noise latent representations from noisy speech $\mathbf{y}$, which can then be used for speech enhancement.

\textbf{Pretrained VAEs}: 
The CVAE and NVAE are independently pretrained in an unsupervised manner using the standard VAE loss, i.e. maximizing the ELBO \cite{Kingma2014},
\begin{align*}
& \mathbb{E}_{q(\mathbf{z}_x|\mathbf{x})}[\log p(\mathbf{x}\vert \mathbf{z}_{x})]\ \hspace{-0.1cm}- \textrm{KL}(q(\mathbf{z}_{x}\vert \mathbf{x})\Vert p(\mathbf{z}_{x})),\tag{3} \label{cvaeloss} \\
& \mathbb{E}_{q(\mathbf{z}_v|\mathbf{v})}[\log p(\mathbf{v}\vert \mathbf{z}_{v})]\ \hspace{-0.1cm}- \textrm{KL}(q(\mathbf{z}_{v}\vert \mathbf{v})\Vert p(\mathbf{z}_{v})),\tag{4} \label{nvaeloss}
\end{align*}
where $\mathbb{E}$ denotes expectation and $\textrm{KL}(\cdot\Vert\cdot)$ denotes the Kullback-Leibler (KL) divergence. In \cite{xiang2022deep}, it is assumed that the posterior distribution and the likelihood for the CVAE follow a multivariate Gaussian distribution with a diagonal covariance matrix, i.e.
\begin{align*}
   & {q({\bf z}_x|{\bf{x}})} = \mathnormal{N} \left({{\bm{\mu}}_{\phi_{x}},\operatorname{diag}(\bm{\sigma}_{\phi_{x}}^2)}\right), \tag{5}\label{cvae_post_distri}\\& p({\bf{x}}|{\bf z}_x) = \mathnormal{N}\left({{\bm{\mu}}_{\theta_{x}},\operatorname{diag}({\bm{\sigma}}_{\theta_{x}}^2)}\right),\tag{6}
\end{align*}
where the mean and variance vectors ${\bm\mu}_{\phi_{x}}$ and ${\bm\sigma}_{\phi_{x}}^2$ are the outputs of the CVAE encoder, and ${\bm\mu}_{\theta_{x}}$ and ${\bm\sigma}_{\theta_{x}}^2$ are the outputs of the CVAE decoder. The prior distribution $p(\mathbf{z}_{x})$ is assumed to be a centered isotropic multivariate Gaussian $p(\mathbf{z}_{x})=\mathnormal{N} ({{\bf{0}},{\bf{I}}})$, where $\bf{I}$ denotes the identity matrix. The reparameterization trick \cite{Kingma2014} is used to sample $\mathbf{z}_x$ from $q(\mathbf{z}_x|\mathbf{x})$ to approximate the intractable expectation in (\ref{cvaeloss}). For the NVAE, similar distributions are assumed as in the CVAE, but these are not explained in detail here.

\textbf{Noisy VAE}: The NSVAE is trained under the supervision of the pretrained CVAE and NVAE. In \cite{xiang2022deep}, it has been proposed to only train the NSVAE encoder and discard the NSVAE decoder. The NSVAE encoder takes the noisy speech $\mathbf{y}$ as input and generates the speech and noise latent representations $\mathbf{z}_x$ and $\mathbf{z}_v$. Aiming at making the posterior distributions $q(\mathbf{z}_x|\mathbf{y})$ and $q(\mathbf{z}_v|\mathbf{y})$ from the NSVAE encoder similar to the posterior distributions $q(\mathbf{z}_x|\mathbf{x})$ and $q(\mathbf{z}_v|\mathbf{v})$ from the pretrained VAEs, the NSVAE is trained by minimizing the loss   
 \begin{align*}
  \textrm{KL}\left({q({\bf z}_x|{\bf{y}})}||{q({\bf z}_x|{\bf{x}})}\right) + \textrm{KL}\left({q({\bf z}_v|{\bf{y}})}||{q({\bf z}_v|{\bf{v}})}\right).
  \tag{7}\label{nsvaeloss}
\end{align*}
Similar to (\ref{cvae_post_distri}), the posterior distributions estimated from the encoder of NSVAE are assumed to follow a multivariate Gaussian distribution, i.e.
\begin{align*}
   & {q({\bf z}_x|{\bf{y}})} = \mathnormal{N} \left({{\bm\mu}_{\phi_{yx}},\operatorname{diag}({\bm\sigma}_{\phi_{yx}}^2)}\right), \tag{8}\\
   & {q({\bf z}_v|{\bf{y}})} = \mathnormal{N}\left({{\bm\mu}_{\phi_{yv}},\operatorname{diag}({\bm\sigma}_{\phi_{yv}}^2)}\right),\tag{9}
\label{nsvaeposterior}
\end{align*} 
where the mean vectors ${\bm\mu}_{\phi_{yx}}$ and ${\bm\mu}_{\phi_{yv}}$, and the variance vectors ${\bm\sigma}_{\phi_{yx}}^2$ and ${\bm\sigma}_{\phi_{yv}}^2$ are the outputs of the NSVAE encoder. The reparameterization trick is also used here to sample speech and noise latent representations $\mathbf{z}_x$ and $\mathbf{z}_v$. Assuming that the sampled latent representations from the posterior distributions $q(\mathbf{z}_x|\mathbf{y})$ and $q(\mathbf{z}_v|\mathbf{y})$ are close to the latent representations from $q(\mathbf{z}_x|\mathbf{x})$ and $q(\mathbf{z}_v|\mathbf{v})$, the sampled latent speech and noise representations from the NSVAE encoder are then used as inputs to the CVAE decoder and the NVAE decoder, respectively. The mean vectors ${\bm{\mu}}_{\theta_{x}}$ and ${\bm{\mu}}_{\theta_{v}}$ from the CVAE and NVAE decoders are used as the estimates of the speech and noise LPS $\hat{\mathbf{x}}$ and $\hat{\mathbf{v}}$. Finally, the clean speech STFT is estimated by applying a real-valued mask to the noisy STFT, i.e.
\begin{equation*} 
    \mathbf{\hat{X}} = \frac{\mathbf{|\hat{X}|}}{\mathbf{\hat{|X|}}+\mathbf{\hat{|V|}}}\mathbf{Y}, \tag{10}
\end{equation*}
where $\mathbf{|\hat{X}|}=10^{\hat{\mathbf{x}}/2}$ and $\mathbf{|\hat{V}|}=10^{\hat{\mathbf{v}}/2}$.

%%%%%%%%%%%%%%%%%%%%%%%%%%%%%%%%%%%%%%%%%%%%%%%%%%%%%%%%%%%%%%%%%%
\section{Loss Function for Pretrained VAEs}
%%%%%%%%%%%%%%%%%%%%%%%%%%%%%%%%%%%%%%%%%%%%%%%%%%%%%%%%%%%%%%%%%%

By training the NSVAE using (\ref{nsvaeloss}), the latent representations $\mathbf{z}_x$ and $\mathbf{z}_v$ from the NSVAE are close to the latent representations $\mathbf{z}_x$ and $\mathbf{z}_v$ from the pretrained VAEs, such that the pretrained VAEs play a crucial role in the complete speech enhancement system. In this paper, we investigate the influence of different latent speech and noise representations of the pretrained VAEs on speech enhancement performance, while keeping the training of the NSVAE unchanged. Instead of training the CVAE and NVAE using the standard VAE loss in (\ref{cvaeloss}) and (\ref{nvaeloss}), we propose to train the CVAE and NVAE using the DIP VAE loss \cite{kumar2018variational}. DIP VAE was proposed to learn disentangled representations, which may offer several advantages such as transferability and interpretability \cite {10579040}. 

Since the pretrained CVAE and NVAE both use the same DIP VAE loss, in the following we will use $\mathbf{z}$ to represent $\mathbf{z}_x$ and $\mathbf{z}_v$, and $\mathbf{s}$ to represent $\mathbf{x}$ and $\mathbf{v}$.
To learn disentangled representations, DIP VAE assumes that all elements of the latent representation $\mathbf{z}$ are uncorrelated with each other. 
This is achieved by matching the covariance of the marginal distribution $q(\mathbf{z})\hspace{-0.08cm}=\hspace{-0.1cm}\int\hspace{-0.05cm} q(\mathbf{z}|\mathbf{s})p(\mathbf{s})d\mathbf{s}$ with the covariance of the prior distribution $p(\mathbf{z})\hspace{-0.07cm}=\hspace{-0.07cm}\mathnormal{N} ({{\bf{0}},{\bf{I}}})$. 
According to the law of total covariances, covariance of $q(\mathbf{z})$ can be written as 
\begin{align}
%\expect_{q_\theta(\vz)} \left[(\vz-\expect_{q_\theta(\vz)}\vz)(\vz-\expect_{q_\theta(\vz)}\vz)^\top\right] = \expect_{p(\vx)} 
\operatorname{Cov}_{q(\mathbf{z})}\hspace{-0.05cm}\left[\mathbf{z}\right] \hspace{-0.05cm} 
&=\hspace{-0.07cm}\mathbb{E}_{q(\mathbf{z})} \hspace{-0.05cm}\left[\left(\mathbf{z}\hspace{-0.05cm}-\hspace{-0.05cm}\mathbb{E}_{q(\mathbf{z})}\hspace{-0.05cm}\left[\mathbf{z}\right]\right)\left(\mathbf{z}\hspace{-0.05cm}-\hspace{-0.05cm}\mathbb{E}_{q(\mathbf{z})}\hspace{-0.05cm}[\mathbf{z}]\right)^{\hspace{-0.05cm}\mathrm{T}}\right]\tag{11}\\
 &=\hspace{-0.05cm}\mathbb{E}_{p(\mathbf{s})} \hspace{-0.03cm}\left(\hspace{-0.03cm}\operatorname{Cov}_{q(\mathbf{z}|\mathbf{s})}\hspace{-0.05cm}[\mathbf{z}]\right) \hspace{-0.05cm}+ \operatorname{Cov}_{p(\mathbf{s})}\hspace{-0.05cm}\left(\mathbb{E}_{q(\mathbf{z}|\mathbf{s})} \hspace{-0.05cm}[\mathbf{z}]\right)\tag{12}\\
 &=\hspace{-0.05cm}\mathbb{E}_{p(\mathbf{s})}\left[\operatorname{diag}(\bm\sigma^2)\right]\hspace{-0.05cm}+ \operatorname{Cov}_{p(\mathbf{s})}\hspace{-0.05cm}\left(\bm\mu\right), \tag{13}
\label{eq:totalcov}
\end{align}
where $\bm\sigma^2$ and $\bm\mu$ are the outputs of the VAE encoder, and $p(\mathbf{s})$ denotes the probability distribution of the input signal (omitted in the following for simplicity). Since $\mathbb{E}
%_{p(\mathbf{s})}%
[\operatorname{diag}(\bm\sigma^2)]$ is a diagonal matrix, the off-diagonal elements in $\operatorname{Cov}_{q(\mathbf{z})}\hspace{-0.05cm}[\mathbf{z}]$ only come from $\operatorname{Cov}
%_{p(\mathbf{s})}%
\hspace{-0.05cm}\left(\bm\mu\right)$.
In \cite{kumar2018variational}, two DIP VAE loss functions were introduced: DIP-VAE-2 aims at matching $\operatorname{Cov}_{q(\mathbf{z})}\hspace{-0.05cm}[\mathbf{z}]$ to the identity matrix, while DIP-VAE-1 neglects $\mathbb{E}
%_{p(\mathbf{s})}%
[\operatorname{diag}(\bm\sigma^2)]$ and aims at matching $\operatorname{Cov}
%_{p(\mathbf{s})}%
\hspace{-0.05cm}\left(\bm\mu\right)$ to the identity matrix \footnote{In this paper, we only consider DIP-VAE-1, since it yielded better results}. This is achieved by penalizing the off-diagonal elements of $\operatorname{Cov}
%_{p(\mathbf{s})}%
\hspace{-0.05cm}\left(\bm\mu\right)$ and encouraging the diagonal elements of $\operatorname{Cov}
%_{p(\mathbf{s})}%
\hspace{-0.05cm}\left(\bm\mu\right)$ to be close to 1,
leading to the disentangling regularizer
\begin{equation}
    \boxed{
    L_{\textrm{reg}}\hspace{-0.05cm}=\hspace{-0.05cm}\lambda_{od}\hspace{-0.05cm}\sum_{i \neq j}\hspace{-0.05cm}\left[\operatorname{Cov}
    %_{p(\mathbf{s})}%
    \hspace{-0.01cm}(\boldsymbol{\mu})\right]_{ij}^2 \nonumber 
	 \hspace{-0.03cm}+\hspace{-0.03cm} \lambda_d \smash{\sum_i}\hspace{-0.03cm}\left( \left[\operatorname{Cov}
     %_{p(\mathbf{s})}%
     \hspace{-0.01cm}(\boldsymbol{\mu})\right]_{ii}\hspace{-0.05cm} -\hspace{-0.05cm} 1 \right)^2\tag{14}\label{reguloss}}
\end{equation}
where $[\cdot]_{ij}$ represents the $(i,j)$-th element of a matrix, and $\lambda_{od}$ and $\lambda_d$ are weighting factors for the off-diagonal elements and the diagonal elements.
Instead of maximizing the ELBO, DIP-VAE is trained by maximizing
\begin{equation}
    \boxed{
    \mathbb{E}_{q(\mathbf{z}|\mathbf{s})}\hspace{-0.03cm}[\log \hspace{-0.01cm} p(\mathbf{s}|\mathbf{z})]
    \hspace{-0.05cm}-\hspace{-0.05cm} \beta \textrm{KL}(q(\mathbf{z}|\mathbf{s}) \Vert p(\mathbf{z}))\hspace{-0.05cm} -\hspace{-0.05cm} L_{\textrm{reg}}}
    \tag{15}\label{dipvaeloss}
\end{equation}
where $\beta$ denotes an additional weighting factor for the KL term. All terms in (\ref{dipvaeloss}), including the KL term, the off-diagonal term and the diagonal term, influence the latent representations of the pretrained VAEs. In the experiments, we will generate different latent representations for the pretrained VAEs by varying $\beta$, $\lambda_{od}$ and $\lambda_d$, and investigate their influence on the speech enhancement performance.  It should be noted that when $\beta=1, \lambda_{od}=0$ and $\lambda_{d}=0$, the loss in (\ref{dipvaeloss}) corresponds to the ELBO for the standard VAE.

%%%%%%%%%%%%%%%%%%%%%%%%%%%%%%%%%%%%%%%%%%%%%%%%%%%%%%%%%%%%%%%%%%
\section{Experiments}
%%%%%%%%%%%%%%%%%%%%%%%%%%%%%%%%%%%%%%%%%%%%%%%%%%%%%%%%%%%%%%%%%%

This section first presents the experimental setup, namely the training and evaluation datasets, the network structure and the training procedure. Then, the experimental results are presented and discussed. 

\subsection{Training and Evaluation Datasets}
To pretrain the CVAE and the NVAE and to train the NSVAE, we used anechoic clean speech and noise from the training set of the DNS3 challenge dataset at a sampling frequency of 16kHz \cite{reddy21_interspeech}. It should be noted that for clean speech we only considered the read speech (leaving out emotional speech), while for noise we did not consider the DEMAND dataset, since it was used for evaluation. We randomly split 50$\%$ of speakers to pretrain the CVAE, 40\% of speakers to train the NSVAE and 10\% of speakers for validation. The noise data was split similarly to pretrain the NVAE and train the NSVAE. To train the NSVAE, the clean speech and noise were randomly mixed using the DNS script at signal-to-noise ratios (SNRs) between -10\,dB and 15\,dB. In total, we generated 30 hours of data for pretraining, 20 hours of data for NSVAE training and 10 hours of data for validation. 

To evaluate the speech enhancement performance, we considered three datasets. As to the matched evaluation dataset, we used the official synthetic DNS3 test set at SNRs between 0\,dB and 19\,dB. To evaluate the generalization ability, we also considered two mismatched datasets with different speakers and noise from the training dataset, namely WSJ0-QUT\cite{bie2022unsupervised} and VoiceBank-DEMAND (VB-DMD)\cite{valentinibotinhao16_interspeech}. WSJ0-QUT contains 1.5 hours of noisy speech, including cafe, home, street and car noise at SNRs of -5\,dB, 0\,dB and 5\,dB. The official VB-DMD test set contains 1 hour of noisy speech, including room, office, bus, cafe and public square noise at SNRs of 2.5\,dB, 7.5\,dB, 12.5\,dB and 17.5\,dB.

\subsection{Network and Training}
In the experiments, we used the same setup for the PVAE system as in \cite{xiang2022deep}. All time-domain signals are transformed to the STFT domain using a Hann window with a frame length of 32\,ms and 50\% overlap. 
The CVAE, NVAE and NSVAE all contain a combination of fully-connected (FC) layers and a uni-directional gated recurrent unit (GRU) layer. The dimension of the latent representation is equal to $L=128$ for all VAEs. The CVAE and NVAE encoders contain three FC layers (ReLU activation) with output dimensions [512, 512, 512], followed by a GRU layer with 512 output units and two parallel FC layers (no activation) to generate the L-dimensional mean and variance vectors: ${\bm\mu}_{\phi_{x}}$ and ${\bm\sigma}_{\phi_{x}}^2$ for the CVAE, and ${\bm\mu}_{\phi_{v}}$ and ${\bm\sigma}_{\phi_{v}}^2$ for the NVAE. The NSVAE encoder has a similar structure as the CVAE and NVAE encoders. The only difference is that the GRU layer is followed by a FC layer with an output dimension of 1024, and four parallel FC layers (no activation) to generate the L-dimensional mean and variance vectors: ${\bm\mu}_{\phi_{yx}}$, ${\bm\mu}_{\phi_{yv}}$, ${\bm\sigma}_{\phi_{yx}}^2$ and ${\bm\sigma}_{\phi_{yv}}^2$.    
The CVAE and NVAE decoders mirror their respective encoders in reverse order, mapping latent representations to LPS. All networks were trained for a maximum of 500 epochs. The training was stopped early in case the validation
loss did not decrease for 20 consecutive epochs. The Adam optimizer with a learning rate of 0.0001 was used. The batch size was set to 128.

\subsection{Experimental Results}
\begin{table*}[!t]	
\setlength{\tabcolsep}{1.7mm}
\centering
\setlength{\abovecaptionskip}{0.05cm}
\caption{{ Average reconstruction SI-SNR (dB) of pretrained CVAE/NVAE (with 95\% confidence interval) on DNS3 dataset, and average SI-SNR (dB) and PESQ (with 95\% confidence interval) of the PVAE system using different loss functions for pretraining and the non-causal RVAE system on different datasets.}}
\label{tab: score_comparison}
\centering

\begin{tabular}{c|c|cc|cc|cc|cc}
	\toprule
	% Add the first column for vertical text
	\multirow{2}{*}{} & \multirow{2}*{Method} & \multicolumn{1}{c}{CVAE} & \multicolumn{1}{c|}{NVAE} & \multicolumn{2}{c|}{DNS3} & \multicolumn{2}{c|}{WSJ0-QUT} & \multicolumn{2}{c}{VB-DMD} \\
	& & \centering SI-SNR & \centering SI-SNR & SI-SNR & PESQ & SI-SNR & PESQ & SI-SNR & PESQ \\
	%\hline
    \cline{2-10}
    &\rule{0pt}{3.5ex}Noisy  & - & - & \makecell{9.10 \\ ($\pm\,$0.88)} & \makecell{1.58 \\ ($\pm\,$0.07)} & \makecell{-2.60 \\ ($\pm\,$0.32)} & \makecell{1.14 \\ ($\pm\,$0.01)} & \makecell{8.40 \\ ($\pm\,$0.38)} & \makecell{1.97 \\ ($\pm\,$0.05)} \\
	\hline \hline
	\multirow{2}{*}{\rotatebox{90}{\scriptsize{With KL\hspace{1mm}}}} &\rule{0pt}{3.5ex}(1) $\beta=1$, $\lambda_{od}=0$, $\lambda_d=0$  & \makecell{14.50 \\ ($\pm\,$0.19)} & \makecell{3.20 \\ ($\pm\,$0.38)} & \makecell{13.10 \\ ($\pm\,$0.75)} & \makecell{2.12 \\ ($\pm\,$0.09)} & \makecell{3.30 \\ ($\pm\,$0.35)} & \makecell{1.35 \\ ($\pm\,$0.02)} & \makecell{13.30 \\ ($\pm\,$0.38)} & \makecell{2.16 \\ ($\pm\,$0.04)}  \\
	%\hline
    \cline{2-10}
    &\rule{0pt}{3.5ex}(2) $\beta=1$, $\lambda_{od}=10^4$, $\lambda_d=10^2$  & \makecell{16.40 \\ ($\pm\,$0.16)} & \makecell{5.10 \\ ($\pm\,$0.34)} & \makecell{13.20 \\ ($\pm\,$0.75)} & \makecell{2.12 \\ ($\pm\,$0.09)} & \makecell{3.10 \\ ($\pm\,$0.47)} & \makecell{1.37 \\ ($\pm\,$0.02)} & \makecell{13.50 \\ ($\pm\,$0.37)} & \makecell{2.13 \\ ($\pm\,$0.04)}  \\
    \hline \hline % This creates a thick line

    \multirow{2}{*}{\rotatebox{90}{\scriptsize{Without KL\hspace{-0.8mm}}}}
    &\rule{0pt}{3.5ex}(3) $\beta=0$, $\lambda_{od}=0$, $\lambda_d=0$  & \makecell{17.10 \\ ($\pm\,$0.23)} & \makecell{3.30 \\ ($\pm\,$0.42)} & \textbf{\makecell{14.10 \\ ($\pm\,$0.70)}} & \textbf{\makecell{2.28 \\ ($\pm\,$0.09)}} & \textbf{\makecell{5.30 \\ ($\pm\,$0.37)}} & \textbf{\makecell{1.50 \\ ($\pm\,$0.03)}} & \makecell{14.80 \\ ($\pm\,$0.33)} & \makecell{2.27 \\ ($\pm\,$0.04)}  \\
    \cline{2-10}
    &\rule{0pt}{3.5ex}(4) $\beta=0$, $\lambda_{od}=10^4$, $\lambda_d=10^2$  & \makecell{16.50 \\ ($\pm\,$0.23)} & \makecell{3.90 \\ ($\pm\,$0.40)} & \makecell{13.80 \\ ($\pm\,$0.71)} & \makecell{2.22 \\ ($\pm\,$0.09)} & \makecell{5.30 \\ ($\pm\,$0.35)} & \makecell{1.49 \\ ($\pm\,$0.02)} & \makecell{16.00 \\ ($\pm\,$0.27)} & \makecell{2.22 \\ ($\pm\,$0.04)}  \\
    \hline \hline
        &\rule{0pt}{3.5ex}Non-causal RVAE\cite{bie2022unsupervised}  & - & - & \makecell{12.40 \\ ($\pm\,$1.30)} & \makecell{1.98 \\ ($\pm\,$0.09)} & \makecell{2.60 \\ ($\pm\,$0.40)} & \makecell{1.33 \\ ($\pm\,$0.02)} & \textbf{\makecell{17.20 \\ ($\pm\,$0.28)}} & \textbf{\makecell{2.41 \\ ($\pm\,$0.04)}} \\
	\bottomrule                             
\end{tabular}	
\label{table1}
\end{table*}

In the experimental evaluation, we investigate the influence of different speech and noise latent representations of the pretrained VAEs on speech enhancement performance. To this end, we consider different values of $\beta$, $\lambda_{od}$ and $\lambda_d$ in the loss (\ref{reguloss}) and (\ref{dipvaeloss}) to pretrain the CVAE and the NVAE. To investigate the influence of the KL term ($\beta$) and the disentangling regularizer ($\lambda_{od}$, $\lambda_d$), we consider the following parameter settings:
\begin{enumerate}[labelindent=1em, leftmargin=1.7em]
    \item Standard VAE\cite{xiang2022deep}: $\beta=1$, $\lambda_{od}=0$, $\lambda_d=0$.
    \item DIP VAE: $\beta=1$, $\lambda_{od}=10^4$, $\lambda_d=10^2$ (for these values of $\lambda_{od}$ and $\lambda_d$, the best speech enhancement performance was obtained on the validation set).
    \item Standard VAE without KL term: $\beta=0$, $\lambda_{od}=0$, $\lambda_d=0$.
    \item DIP VAE without KL term: $\beta=0$, $\lambda_{od}=10^4$, $\lambda_d=10^2$.
\end{enumerate}
In addition to investigating the influence of these parameters on the performance of the causal PVAE system, we also include the non-causal unsupervised recurrent VAE (RVAE) proposed in \cite{bie2022unsupervised} as an additional VAE-based speech enhancement system. The RVAE was trained on the same clean speech dataset as the CVAE. For evaluation metrics, we considered the Scale-Invariant Signal-to-Noise Ratio (SI-SNR) and the wide-band Perceptual Evaluation of Speech Quality (PESQ)\cite{941023}, using clean speech signal as the reference signal.

For all considered VAE-based speech enhancement systems, Table \ref{table1} shows the speech enhancement performance in terms of average SI-SNR and PESQ (with 95\% confidence interval) for the matched dataset (DNS3) and for both mismatched datasets (WSJ0-QUT, VB-DMD). In addition, Table \ref{table1} also shows the reconstruction ability of the pretrained VAEs in terms of the average SI-SNRs (with 95\% confidence interval) for the CVAE and NVAE on the DNS3 dataset. In general, it can be observed that the reconstruction ability of the CVAE is better than the reconstruction ability of the NVAE. This can be explained by the fact that clean speech spectrograms exhibit more regular harmonic structures and temporal patterns than noise spectrograms, making them easier for VAE to model and reconstruct.
It can be observed that the best reconstruction ability for the CVAE is obtained by the setting (3), while the best reconstruction ability for the NVAE is obtained by the setting (2).

For the matched dataset, DNS3 dataset, it can be observed that for all considered parameter settings the causal PVAE system outperforms the non-causal RVAE system in terms of SI-SNR and PESQ. A relationship between the speech enhancement performance of the PVAE system and the reconstruction ability of the CVAE can be observed. It can also be observed that the disentangling regularizer in (\ref{dipvaeloss}) is not able to significantly improve the speech enhancement performance, i.e. setting (2) only yields an SI-SNR improvement of 0.1 compared to the standard VAE setting (1). When comparing setting (3) to setting (1) and setting (4) to setting (2), it can be observed that the KL term has a large influence on the speech enhancement performance. More in particular, omitting the KL term in the loss (\ref{dipvaeloss}) for pretraining CVAE and NVAE appears to be advantageous to speech enhancement performance. The best speech enhancement performance is obtained in setting (3), yielding an SI-SNR improvement of 1.0 and a PESQ improvement of 0.16 compared to the standard VAE setting (1) and an SI-SNR improvement of 1.7 and a PESQ improvement of 0.3 compared to the RVAE system.  

\begin{figure}[!t]
	\centering
	
	\begin{minipage}{0.23\textwidth}
		\centering
		\includegraphics[width=\linewidth]{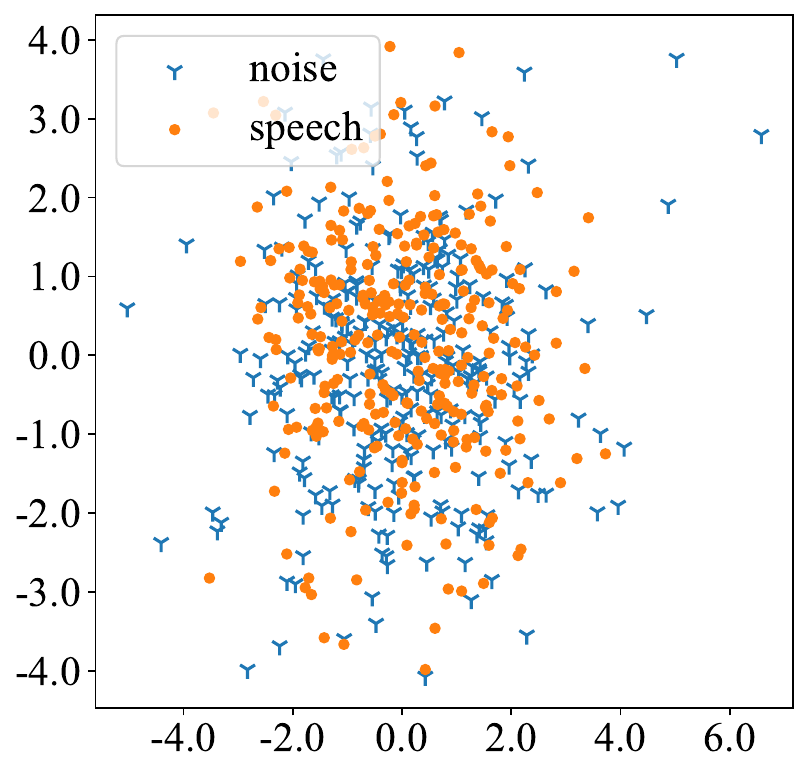} % Replace with your image file
        \vspace{-0.7cm}
        \captionsetup{justification=centering} % Ensure caption is centered
		\caption*{\resizebox{0.78\linewidth}{!}{(1) $\beta=1$, $\lambda_{od}=0$, $\lambda_d=0$}} % Subtitle for the first image
	\end{minipage} \hfill
	\begin{minipage}{0.23\textwidth}
		\centering
		\includegraphics[width=\linewidth]{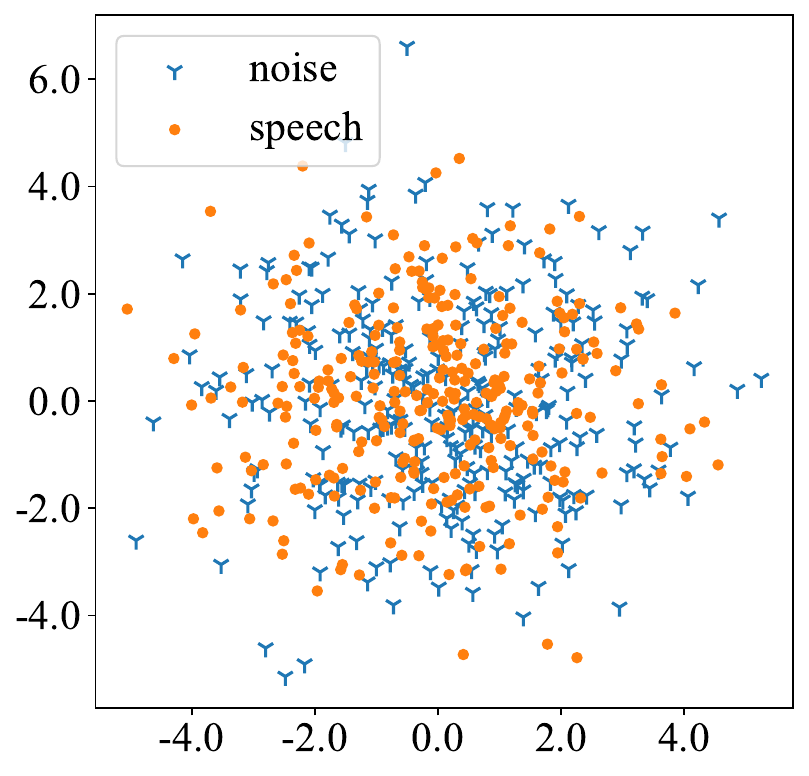} % Replace with your image file
        \vspace{-0.7cm}
        \captionsetup{justification=centering} % Ensure caption is centered
		\caption*{\resizebox{0.9\linewidth}{!}{(2) $\beta=1$, $\lambda_{od}=10^4$, $\lambda_d=10^2$}} % Subtitle for the second image
	\end{minipage} \\
	
	% \vspace{0.2cm} % Add some vertical space between the rows
	
	\begin{minipage}{0.23\textwidth}
		\centering
		\includegraphics[width=\linewidth]{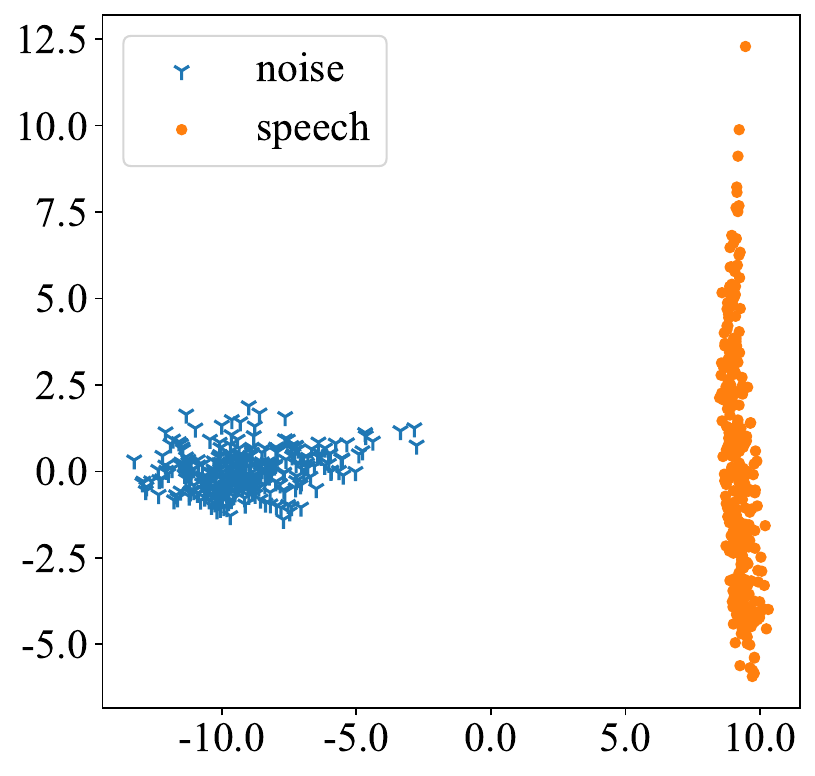} % Replace with your image file
        \vspace{-0.7cm}
        \captionsetup{justification=centering} % Ensure caption is centered
		\caption*{\resizebox{0.78\linewidth}{!}{
        (3) $\beta=0$, $\lambda_{od}=0$, $\lambda_d=0$}} % Subtitle for the third image
	\end{minipage} \hfill
	\begin{minipage}{0.23\textwidth}
		\centering
		\includegraphics[width=\linewidth]{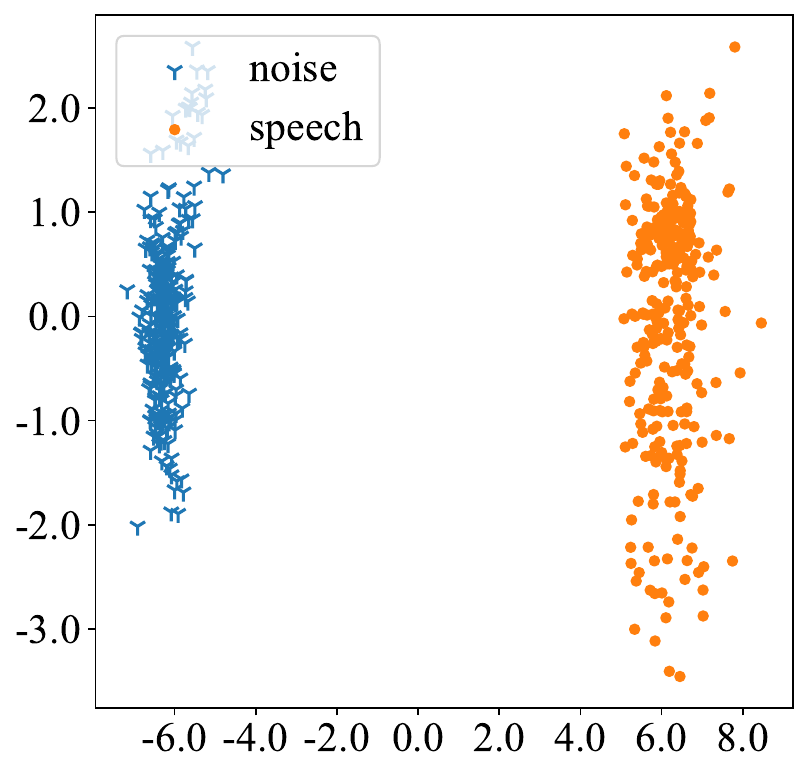} % Replace with your image file
            \vspace{-0.7cm}
            \captionsetup{justification=centering} % Ensure caption is centered
		\caption*{\resizebox{0.9\linewidth}{!}{(4) $\beta=0$, $\lambda_{od}=10^4$, $\lambda_d=10^2$}} % Subtitle for the fourth image
	\end{minipage}
	
	\caption{Latent speech and noise representations from the NSVAE encoder evaluated on the DNS3 dataset using different loss functions for pretraining the CVAE and the NVAE.} % Title for the whole figure
	\label{latentfig} % Label for referencing the figure
\end{figure}

To better understand the influence of the KL term, Fig. \ref{latentfig} depicts the speech and noise latent representations, $\mathbf{z}_x$ and $\mathbf{z}_v$, generated by the NSVAE encoder on the DNS3 dataset for all considered parameter settings. For visualization purposes, we used Principle Component Analysis\cite{10.1145/3447755} to project the high-dimensional latent representations to a 2-dimensional latent space. It should be noted that we decided to depict the latent representations from the NSVAE encoder and not from the pretrained CVAE and NVAE encoders, since 1) the NSVAE encoder is used for speech enhancement, and 2) the latent representations from the NSVAE and the CVAE and NVAE are similar (not shown here). In the top figures ($\beta=1$), it can be observed that the speech and noise latent representations both cluster around the origin of the latent space. This can be explained by the fact that the KL term in (\ref{dipvaeloss}) favors latent representations to have the same distribution $p(\mathbf{z})=\mathnormal{N}(\mathbf{0},\mathbf{I})$ in the latent space. In the bottom figures ($\beta=0$), it can be observed that omitting the KL term clearly separates the latent representations of clean speech and noise. In this setting, the pretrained models are free to use the latent dimensions to encode detailed information without being penalized by the KL term. This suggests that the separation of speech and noise in the latent space contributes to improved speech enhancement performance of the PVAE system. 

For mismatched datasets, similar findings can be observed as for the matched dataset. For the WSJ0-QUT dataset, which has a much lower SNR range than the DNS3 dataset, the causal PVAE system outperforms the non-causal RVAE system for all considered parameter settings, with setting (3) yielding the best speech enhancement performance in terms of SI-SNR and PESQ. For the VB-DMD dataset, the best speech enhancement is obtained by the non-causal RVAE system. Nevertheless, among the PVAE systems, the best performance is obtained for parameter settings that omit the KL term for pretraining, i.e. setting (4) in terms of SI-SNR and setting (3) in terms of PESQ.

%%%%%%%%%%%%%%%%%%%%%%%%%%%%%%%%%%%%%%%%%%%%%%%%%%%%%%%%%%%%%%%%%%
\section{Conclusions}
%%%%%%%%%%%%%%%%%%%%%%%%%%%%%%%%%%%%%%%%%%%%%%%%%%%%%%%%%%%%%%%%%%
In this paper, we investigated the influence of different speech and noise latent representations of pretrained VAEs on the speech enhancement performance of the PVAE system. More in particular, we explored how the different terms in the DIP VAE loss affect the latent representations and hence the speech enhancement performance. 
Experimental results on several datasets demonstrate that omitting the KL term to pretrain the CVAE and the NVAE significantly improves speech enhancement performance compared to the standard VAE. This separates the clean speech and noise representations in the latent space, whereas the standard VAE causes these representations to overlap. 
In future work, similar investigations will be conducted on complex-valued VAE-based speech enhancement systems.

%%%%%%%%%%%%%%%%%%%%%%%%%%%%%%%%%%%%%%%%%%%%%%%%%%%%%%%%%%%%%%%%%%
% BIBLIOGRAPHY
%%%%%%%%%%%%%%%%%%%%%%%%%%%%%%%%%%%%%%%%%%%%%%%%%%%%%%%%%%%%%%%%%%
\small
\bibliographystyle{ieeetr}
\bibliography{example}

%%%%%%%%%%%%%%%%%%%%%%%%%%%%%%%%%%%%%%%%%%%%%%%%%%%%%%%%%%%%%%%%%%

\end{document}